\documentclass[12pt]{article}
\usepackage{graphicx}
\usepackage{dcolumn}
\usepackage{bm}
\usepackage{amssymb}
\usepackage{multirow}

\newcommand\pubdate{\today}

\newcommand\pubnumber{FERMILAB-CONF-11-284-E}

\def\Title#1{\begin{center} {\Large #1 } \end{center}}
\def\Author#1{\begin{center}{ \sc #1} \end{center}}
\def\Address#1{\begin{center}{ \it #1} \end{center}}

\newcommand\pubblock{\rightline{\begin{tabular}{l} \pubnumber\\
         \pubdate  \end{tabular}}}
\newenvironment{Abstract}{\begin{center}{\bf Abstract}\end{center} \bigskip \begin{quotation}  }{\end{quotation}}
\newenvironment{Presented}{\begin{quotation} \begin{center} 
             PRESENTED AT\end{center}\bigskip 
      \begin{center}\begin{large}}{\end{large}\end{center} \end{quotation}}

\def\beq{\begin{equation}}
\def\eeq#1{\label{#1}\end{equation}}
\def\eeqn{\end{equation}}

\def\beqa{\begin{eqnarray}}
\def\eeqa#1{\label{#1}\end{eqnarray}}
\def\eeqan{\end{eqnarray}}

\let\bar=\overbar

\def\Dslash{\not{\hbox{\kern-4pt $D$}}}
\def\dslash{\not{\hbox{\kern-2pt $\del$}}}

\def\msb{{\bar{\ssstyle M \kern -1pt S}}}

\newcommand {\Bs} {\ensuremath{B^0_s}}
\newcommand {\Bq} {\ensuremath{B^0_q}}
\newcommand {\Ds} {\ensuremath{D_s}}

\newcommand {\asld} {\ensuremath{a^d_{\mathrm{sl}}}}
\newcommand {\asls} {\ensuremath{a^s_{\mathrm{sl}}}}
\newcommand {\aslq} {\ensuremath{a^q_{\mathrm{sl}}}}
\newcommand {\aslb} {\ensuremath{A^b_{\mathrm{sl}}}}

\newcommand{\AmS}{{\protect\the\textfont2
  A\kern-.1667em\lower.5ex\hbox{M}\kern-.125emS}}

\begin{document}
\begin{titlepage}
\pubblock

\vfill


\Title{Evidence for an anomalous like-sign dimuon charge asymmetry}
\vfill
\Author{M.~R.~J.~Williams, on behalf of the D0 Collaboration}
\Address{Department of Physics, Lancaster University, \\ 
        Bailrigg, Lancashire LA1 4YB, U.K.}
\vfill


\begin{Abstract}
We present a measurement of the like-sign dimuon asymmetry in semileptonic $b$-hadron decays, performed using 6.1~fb$^{-1}$ of $p\bar{p}$ collisions recorded with the D0 detector at a center-of-mass energy $\sqrt{s}=1.96$~TeV at the Fermilab Tevatron collider. This measured value is $\aslb=[-0.957\pm0.251\thinspace({\rm stat})\pm0.146\thinspace({\rm syst})]$~\%, which disagrees with the Standard Model prediction at a statistical level of 3.2~$\sigma$, and provides the first evidence of anomalous $CP$ violation in the mixing of neutral $B$ mesons.
\end{Abstract}

\vfill

\begin{Presented}
The Ninth International Conference on\\
Flavor Physics and CP Violation\\
(FPCP 2011)\\
Maale Hachamisha, Israel,  May 23--27, 2011
\end{Presented}
\vfill

\end{titlepage}
\def\thefootnote{\fnsymbol{footnote}}
\setcounter{footnote}{0}
%


\section{Introduction}

During the big-bang origin of the universe, fundamental symmetry principles imply that particles and antiparticles were created in equal quantities. However, current observations show the universe to be matter dominated. Baryogenesis provides a mechanism to generate the observed asymmetry and requires a number of conditions, one of which is the violation of $CP$ symmetry~\cite{sakharov}. The Standard Model (SM) naturally includes $CP$ violation $(CPV)$ through complex phases induced by quark mixing. While measurements in the $K^0$ and $B^0_d$ systems show excellent agreement with the SM predictions~\cite{pdg}, the observed level of $CP$ violation is far too small to yield the observed matter-antimatter asymmetry~\cite{gavela_huet}. As such it is important to search for new sources of $CPV$ beyond the SM. The $B^0_s$ system is one possible source, since the effects of $CPV$ are expected to be small in the SM~\cite{Nierste}, and can be significantly enhanced by new physics models~\cite{Randall,Hewett,Hou,Soni,Buras}. 

$CPV$ effects can be induced via the mixing process whereby a neutral $B$ meson oscillates into its antiparticle $\bar{B}$. In this case, the experimental observable is the asymmetry between the processes $B_{s,d}^0 \to \bar{B}_{s,d}^0 \to \bar{f}$ and $\bar{B}_{s,d}^0 \to B_{s,d}^0 \to f$, where $f$ is the final state decay product. In particular, for semileptonic decays to a muonic final state, the flavor specific asymmetry is defined as:
\begin{eqnarray}
\label{eq:aqsl}
a^q_{\rm sl} \equiv \frac{\Gamma(\bar B_q \to B_q \to \mu^+ X) - \Gamma(B_q \to \bar B_q \to \mu^- X)}
             {\Gamma(\bar B_q \to B_q \to \mu^+ X) + \Gamma(B_q \to \bar B_q \to \mu^- X)}. \nonumber 
\end{eqnarray}
For the case where the original flavor ($B^0_d$, $B^0_s$) is not distinguished, the asymmetry has contributions from both systems, and is denoted $a^b_{sl} =  \beta_d \asld + \beta_s \asls$.  At the Tevatron accelerator, the coefficients have been measured as $\beta_d = 0.506 \pm 0.043$ and $\beta_s = 0.494 \pm 0.043$~\cite{pdg}. The asymmetries are related to the meson mixing parameters by the relation:
\begin{eqnarray}
\aslq & = & \frac{\Delta \Gamma_q}{\Delta M_q} \tan \phi_q, \label{i_phiq}
\end{eqnarray}
where $\phi_q$ is the $CP$-violating phase, and $\Delta M_q$ ($\Delta \Gamma_q$) is the mass (width) difference between the eigenstates of the neutral $\Bq$ meson systems. The SM predicts $\asld = (-4.8 ^{+1.0}_{-1.2}) \times 10^{-4}$, $\asls = (2.06 \pm 0.57) \times 10^{-5}$~\cite{Nierste}, and hence:
\begin{equation}
a^b_{sl}({\rm SM}) = [-0.023^{+0.005}_{-0.006}]~\%.
\label{in_aslbsm}
\end{equation}

Experimentally, it is a challenge to measure such an asymmetry to high precision, in part because of dilution from background processes where the mesons don't mix prior to decay. A more sensitive parameter is the like-sign dimuon asymmetry from semileptonic decays:
\begin{equation}
\aslb \equiv \frac{N^{++}_{b} - N^{--}_{b}}{N^{++}_{b} + N^{--}_{b}},
\end{equation}
where $N^{++(--)}_{b}$ are the number of events containing two $b$-hadrons decaying semileptonically into two positive (negative) muons. For an original $b\bar{b}$ state to produce like-sign dimuons in the final state, one of the quarks must have hadronised to a neutral meson, and then mixed prior to decay, while the other must decay without mixing. 
In fact, there is a second source of such muons: sequential semileptonic decays $b \rightarrow c\mu^- X \rightarrow \mu^+\mu^+ X^{\prime}$. Such events are suppressed in this analysis by the requirement $M(\mu\mu) > 2.8$~GeV, and their contributions are taken into account. 
The fundamental asymmetry being probed is unchanged by requiring a second muon, i.e. $\aslb = a^b_{sl}$.

In this document, we present a measurement of $\aslb$, which takes advantage of the two different ways it can be extracted to minimise the total experimental uncertainty on the asymmetry. This study was performed using $6.1$ fb$^{-1}$ of data recorded with the D0 detector~\cite{d0det} at the Fermilab Tevatron proton-antiproton ($p\bar{p}$) collider, operating at a center-of-mass energy of 1.96~TeV. More detailed information can be found in the official publication~\cite{d0asym2mu}. 

The D0 experiment is well suited to the investigation of the small effects of $CP$ violation because the periodic reversal of the D0 solenoid and toroid magnetic field polarities results in a cancellation of most detector-related asymmetries. In addition, the $p\bar p$ initial state is a $CP$ eigenstate, and the high center-of-mass energy provides access to mass states beyond the reach of the $B$-factories running at $\sqrt{s}=M(\Upsilon(4S))$.

\section{Measurement Overview}

To extract $\aslb$, the raw asymmetries must first be measured. The like-sign dimuon charge asymmetry $A$ is defined as
\begin{equation}
A \equiv \frac{N^{++} - N^{--}}{N^{++} + N^{--}},
\label{o_defA}
\end{equation}
where $N^{++(--)}$ is the number of events containing two positive (negative) muons which fulfill the kinematic requirements (see below). This quantity is related to $\aslb$ by:
\begin{equation}
A = K \cdot \aslb + A_{\mathrm{bkg}}.
\label{eq:A_to_Absl}
\end{equation}
Here $A_{\mathrm{bkg}}$ represents the contribution to the asymmetry from background processes, defined as those in which the muon arises from the decay of a long-lived particle, and where detector asymmetries can therefore play a role. The coefficient $K$ accounts for dilution from signal-like events (i.e. weak decays of heavy quarks) without $B^0_q$ mixing, which only contribute to the denominator in Eq.~(\ref{o_defA}). 

Similarly, the inclusive muon asymmetry is defined as 
\begin{eqnarray}
a \equiv \frac{n^{+} - n^{-}}{n^{+} + n^{-}} = k \cdot \aslb + a_{\mathrm{bkg}},
\label{o_defa}
\end{eqnarray}
with the lower-case symbols representing the equivalent inclusive muon parameters as those described for the dimuon case above.

The general strategy of the measurement is as follows. The raw asymmetries $A$ and $a$ are first obtained by event counting, following application of all selection criteria. The parameters $A_{\mathrm{bkg}}$, $a_{\mathrm{bkg}}$, $K$ and $k$ are determined in bins of muon transverse momenta, using data-driven methods wherever possible. Equations~(\ref{o_defA}--\ref{o_defa}) are then used to extract $\aslb$ separately for the inclusive and dimuon cases. Finally, the two measurements are combined, taking advantage of correlations in their uncertainties, to yield a single measurement of $\aslb$ with significantly improved precision.

We define all muons from weak decays of $b$ and $c$ quarks as signal, and all other muons (dominated by pion and kaon decay-in-flight) as background. The dilution coefficients are determined from Monte Carlo simulation, using the branching fractions and momentum spectra of signal processes to evaluate their relative contributions to the signal sample. Apart from the asymmetry-inducing semileptonic $B$ decays, contributions from the following processes are taken into account in the simulation: sequential decays $b \rightarrow c \rightarrow \mu^+ X$, decays $b \to c\bar{c}q$, $c\bar{c}$ and $b\bar{b}c\bar{c}$ production, and the dimuon decays of narrow resonances, such as $\eta$, $\omega$, $\rho^0$, $\phi(1020)$ and J/$\psi$. The obtained values are $k = 0.070 \pm 0.006$ and $K = 0.486 \pm 0.032$, showing that the raw dimuon asymmetry is more sensitive to $\aslb$ than the inclusive muon asymmetry, as expected from the reduction in background contributions.

The largest background asymmetry arises from the fact that $K^+$ and $K^-$ mesons interact differently with the material of the detector (which is clearly matter-antimatter asymmetric), and thus their decay rates into positive and negative muons are not identical. Background contributions from kaon and pion decay-in-flight, hadronic punchthrough to the muon system, and residual reconstruction asymmetries, are all measured using collision data.

\section{Event Selection}

The inclusive muon and like-sign dimuon samples are collected from events satisfying single and dimuon triggers, respectively.  Muon candidates are defined by matching central tracks with a `local muon' track reconstructed in the muon system and by applying tight quality requirements aimed at reducing backgrounds such as cosmic rays, beam halo, and mis-matched muon-track combinations. All muons must have transverse momentum satisfying $ 1.5 < p_T < 25$~GeV, and lie in the geometrical region corresponding to pseudorapidity $|\eta| < 2.2$. For muons with $p_T < 4.2$~GeV, an additional requirement is applied to the longitudinal component $|p_z| > 6.4$~GeV, to ensure that the muon has sufficient penetrating power to reach the outermost layers of the muon system. The transverse impact parameter of the muon track relative to the reconstructed primary $p\bar p$ interaction vertex (PV) must be smaller than $0.3$~cm, and this point-of-closest-approach must lie within $0.5$ cm of the PV along the beam axis.  
Muons fulfilling the above requirements define the inclusive muon sample, used to determine $a$ in Eq.~\ref{o_defa}, with no constraint on the number of muons per event. The dimuon sample contains all events with at least two muon candidates of the same charge. To reduce the contribution of background from sequential decays $B \to \mu DX \to \mu\mu X^{\prime}$, the two muons must have an invariant mass above 2.8~GeV.

\section{Detector Asymmetries}

The background asymmetries $a_{\mathrm{bkg}}$ and $A_{\mathrm{bkg}}$ comprise contributions from several sources. These include secondary muons from decays-in-flight of charged kaons and pions; fake muon signals from hadronic `punch-through' of kaons, pions, and protons to the muon system; and residual detector asymmetries. For example, the inclusive muon background asymmetry is expressed as:
\begin{eqnarray}
a_{\mathrm{bkg}} = f_K \cdot a_K + f_{\pi} \cdot a_{\pi} + f_p \cdot a_p \nonumber \\
+ (1 - f_K - f_{\pi} - f_p) \cdot \delta.
\label{abkg_def}
\end{eqnarray}
Here, the first three terms account for asymmetries from kaons, pions, and protons, respectively. In each case, the total asymmetry is the product of the fraction $f_i$ of muons arising from this source, and the asymmetry $a_i$ associated with that source. The last term accounts for any remaining detector asymmetry, $\delta$, on the signal muon sample. The dominant source of asymmetry is the kaon term, since the $K^-$ meson has a significantly shorter interaction length in matter than the $K^+$ meson~\cite{pdg}, and therefore has less chance of decaying to a muon. The resulting contribution to the raw asymmetry $a$ is therefore positive, and at least a factor ten larger than any of the other three terms in Eq.~(\ref{abkg_def}). 

\begin{figure}[!hb]
\begin{center}
\includegraphics[width=0.7\textwidth]{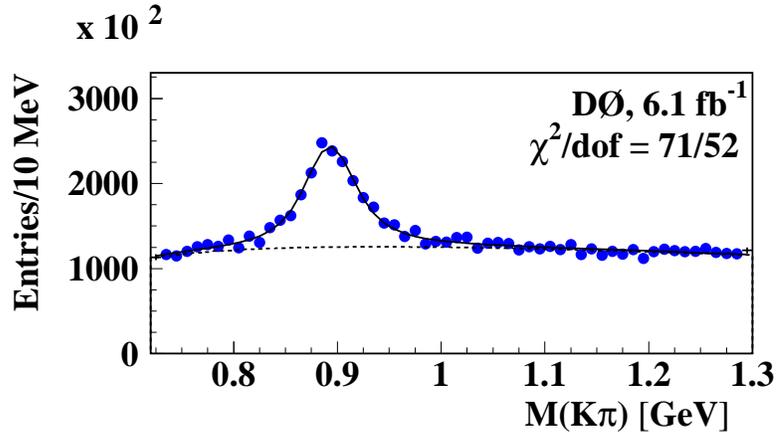}
\caption{The difference in the number of events for the $K^+ \pi^-$ and $K^- \pi^+$ mass distributions of $K^{*0}$
candidates in the inclusive muon sample. The solid line represents the result of the fit, while the dashed line
shows the background contribution.}
\label{fig_kst0_diff}
\end{center}
\end{figure}

The asymmetry from kaons $a_K$ is measured in data using $K^{*0} \to K^{\pm} \pi^{\mp}$ decays. The $K^{*0}$ invariant mass distribution is constructed by assigning the charged kaon mass to all muons in the inclusive muon sample and combining with another track (assigned the pion mass). The difference of distributions for events with $K^+ \to \mu^+$ and $K^- \to \mu^-$ tracks is shown in Fig.~\ref{fig_kst0_diff}, demonstrating the large excess of positive $K \to \mu$ candidates. Simulations are used to correct for the fraction of kaons ($\approx6\%$) that decay prior to being reconstructed~\cite{d0asym2mu}. The measurement is repeated in an independent sample of kaons from $\phi \to K^+ K^-$ decays, and the results are fully consistent. Similarly, the asymmetry for pion or proton tracks misidentified as muons ($a_{\pi}$, $a_p$) are measured using samples of $K_S \to \pi^+ \pi^-$ and $\Lambda \to p \pi^-$ decays, respectively.

The fractions of muons from kaons, $f_K$, is extracted using a similar method, under certain theoretical assumptions which are validated in data. Additional information from both data and simulation is then used to relate the kaon fraction to the corresponding pion and proton fractions $f_{\pi}$ and $f_p$.

After measuring the fractions of muons originating from kaons, pions and protons, the remaining signal (heavy flavor) muons constitute $[58.1 \pm 1.4~({\rm stat}) \pm 3.9~({\rm syst})]$\% of the inclusive muon sample. For the like-sign dimuon case, a fraction $[66.5 \pm 1.6~({\rm stat}) \pm 3.3~({\rm syst})]$\% of the candiate events have both muons produced from heavy flavor decays.

The effects of detector asymmetries on this measurement are significantly suppressed by the regular reversal of the solenoid and toroid polarities in the D0 detector. Using $J/\psi \to \mu^+\mu^-$ decays, the residual reconstruction asymmetry is measured as a function of muon $p_T$, and found to be of order $10^{-3}$. More detail on such effects can be found in Ref.~\cite{D01}.

\begin{table}
\caption{Sources of uncertainty on $\aslb$ in Eqs.~(\ref{ah1},\ref{ah2},\ref{ah3}). The first eight rows correspond to statistical uncertainties and the next three rows to systematic uncertainties.}\label{tab5}
\centering
\begin{tabular}{@{}|c|ccc|}
\hline
Source & $\delta(\aslb)$ & $\delta(\aslb)$ & $\delta(\aslb)$ \\
       &   (\ref{ah1})   &   (\ref{ah2})   &  (\ref{ah3})  \\
\hline
$A$ or $a$ (stat)       & 0.00066 & 0.00159 & 0.00179 \\
$K$ fraction (stat)     & 0.00222 & 0.00123 & 0.00140 \\
$\pi$ fraction          & 0.00234 & 0.00038 & 0.00010 \\
$p$ fraction            & 0.00301 & 0.00044 & 0.00011 \\
$K$ asym.               & 0.00410 & 0.00076 & 0.00061 \\
$\pi$ asym.             & 0.00699 & 0.00086 & 0.00035 \\
$p$ asym.               & 0.00478 & 0.00054 & 0.00001 \\
detector asym.          & 0.00405 & 0.00105 & 0.00077 \\
\hline
$K$ fraction            & 0.02137 & 0.00300 & 0.00128 \\
$\pi$, $K$, $p$         & \multirow{2}{*}{0.00098} & \multirow{2}{*}{0.00025} & \multirow{2}{*}{0.00018} \\
multiplicity            &         &         &         \\
$\aslb$ dilution        & 0.00080 & 0.00046 & 0.00068 \\
\hline
Total statistical       & 0.01118 & 0.00266 & 0.00251 \\
Total systematic        & 0.02140 & 0.00305 & 0.00146 \\
Total                   & 0.02415 & 0.00405 & 0.00290 \\
\hline
\end{tabular}
\end{table}

\section{Results}

The uncorrected asymmetries $a$ and $A$ are obtained by counting events in the inclusive muon sample (containing $1.495 \times 10^9$ muons) and like-sign dimuon sample (containing $3.731 \times 10^6$ events), respectively, and measured to be:
\begin{eqnarray}
\label{value_a}
a & = & [+0.955 \pm 0.003~({\rm stat})]~\%, \\
A & = & [+0.564 \pm 0.053~({\rm stat})]~\%.
\label{value_A}
\end{eqnarray}
After correcting for background ($a_{bkg}$, $A_{bkg}$) and for the dilutions ($k$, $K$) of the signal component, we obtain:
\begin{equation}
\aslb = [ +0.94 \pm 1.12~({\rm stat}) \pm 2.14~({\rm syst})]~\%
\label{ah1} 
\end{equation}
from the inclusive muon sample and
\begin{eqnarray}
\aslb = [ -0.736 \pm 0.266~({\rm stat}) \pm 0.305~({\rm syst}) ]~\%
\label{ah2}
\end{eqnarray}
from the like-sign dimuon sample. The uncertainties on the latter measurement are much smaller, due to the difference in the dilution coefficients $k \ll K$ in Eqs.~(\ref{eq:A_to_Absl}--\ref{o_defa}). Since the same background processes contribute to the uncorrected asymmetries $a$ and $A$, their uncertainties are strongly correlated. The systematic uncertainties can therefore be significantly reduced by suitable linear combination of the raw asymmetries:
\begin{equation}
A' \equiv A - \alpha a.
\label{combination}
\end{equation}
The total uncertainty is minimised by scanning the coefficient $\alpha$ over a wide range, yielding the optimal value $\alpha=0.959$. The final result for the asymmetry $\aslb$ is then:
\begin{equation}
\label{ah3}
\aslb = [-0.957 \pm 0.251~({\rm stat}) \pm 0.146~({\rm syst})]~\%.
\end{equation}
It differs by 3.2 standard deviations from the SM prediction for $\aslb$ of Eq.~(\ref{in_aslbsm}). The contributions to the total uncertainty of $\aslb$ in Eqs.~(\ref{ah1},\ref{ah2},\ref{ah3}) are listed in Table~\ref{tab5}, showing that the combination reduces the systematic effects to the point where the precision is dominated by statistical uncertainties.

\section{Cross-Checks}

\begin{figure}
\begin{center}
\includegraphics[width=0.7\textwidth]{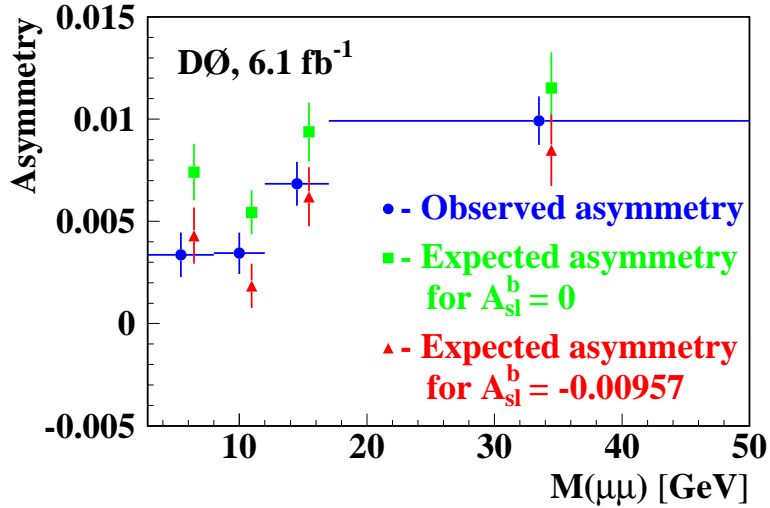}
\caption{ The observed and expected like-sign dimuon charge asymmetry $A$ as a function of dimuon invariant mass. The expected asymmetry is shown for $\mbox{\aslb=0}$ and $\mbox{\aslb=-0.00957}$.}
\label{a-mmm}
\end{center}
\end{figure}

To test the stability of the measurement under a range of different event selection criteria, several consistency checks are performed. In each case, the procedure to extract $\aslb$ is repeated over a sub-sample of data, defined by applying new criteria on detector era, muon and track quality requirements, impact parameter, transverse momentum, polar angle and pseudorapidity. While the raw asymmetries $a$ and $A$ vary significantly, by up to 140\% (as a result of different sample composition), the obtained variations of $\aslb$ are consistent with purely statistical fluctuations with respect to the nominal value.

The small value of the dilution coefficient $k$ in Eq.~(\ref{eq:A_to_Absl}) shows that the raw inclusive muon asymmetry is dominated by background contributions. This is verified by  comparing the observed asymmetry with the background expectation $a_{bkg}$ as a function of muon $p_T$, and excellent agreement is found. In addition, the expected invariant mass dependence of the like-sign dimuon asymmetry is reproduced in data, provided that $\aslb$ is fixed to its measured value. The alternative hypothesis $\aslb=0$ leads to significant discrepancy, as shown in Fig.~\ref{a-mmm}.

\begin{figure}[!hb]
\begin{center}
\includegraphics[width=0.7\textwidth]{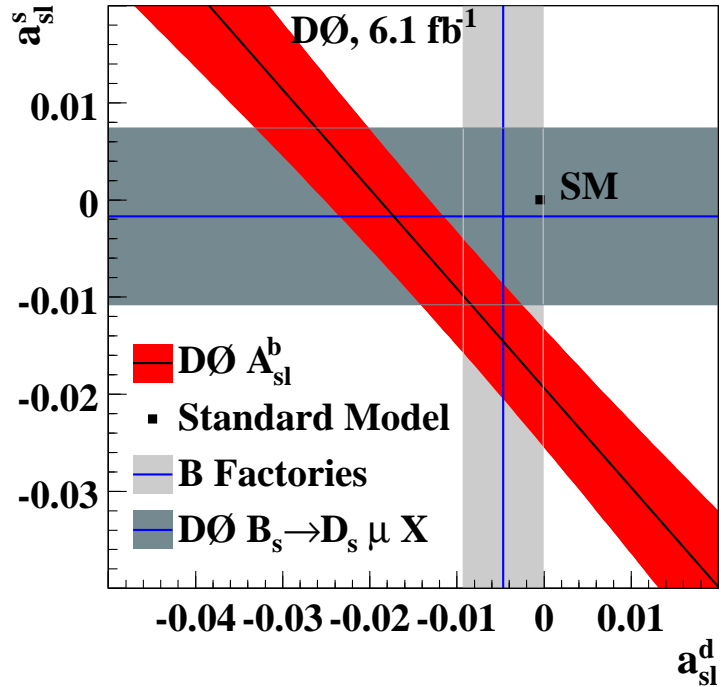}
\caption{Comparison of $\aslb$ in data with the SM prediction
for $\asld$ and $\asls$. Also shown are other measurements of 
\mbox{$\asld=-0.0047\pm0.0046$}~\cite{babar,belle,hfag} and
\mbox{$\asls=-0.0017\pm0.0091$}~\cite{asl-d0}. The bands 
represent the $\pm 1$ standard deviation uncertainties on
each measurement.}
\label{asl_svsd}
\end{center}
\end{figure}

\section{Interpretations and Conclusions}

The asymmetry $\aslb$ is a linear combination of the flavor-specific asymmetries $a^d_{sl}$ and $a^s_{sl}$, and therefore places
constraints on the allowed phase space in the $\asld$--$\asls$ plane, as shown in Fig.~\ref{asl_svsd}. Shown on the same axis are independent constraints on $\asld$ (from the $B$-factories~\cite{babar,belle,hfag}) and on $\asls$ (from the D0 measurement in $\Bs \to \Ds \mu X$ decays~\cite{asl-d0}). Using additional constraints on $a^d_{sl}$ allows the $CP$-violating phase $\phi_s$ of the $\Bs$ mixing matrix to be constrained, as shown in Eq.(~\ref{i_phiq}). This extraction, and comparisons with related measurements, are given in Ref.~\cite{d0asym2mu}.

In conclusion, using a sample of $p\bar{p}$ collison data, corresponding to 6.1 fb$^{-1}$ of integrated luminosity, we have measured the like-sign dimuon charge asymmetry $\aslb$ of semileptonic $b$-hadron decays:
\begin{equation}
\aslb = [-0.957\pm0.251\thinspace({\rm stat})\pm0.146\thinspace({\rm syst})]~\%.
\label{Ab_6}
\end{equation}
This measurement is consistent with, and supersedes, our previous measurement\cite{D01} obtained with 1~fb$^{-1}$.
The asymmetry disagrees with the prediction of the SM by 3.2 standard deviations, and gives the first evidence for anomalous $CP$ violation in the mixing of neutral $B$ mesons.

\section{Future Plans}

The D0 collaboration are working on an updated measurement of the asymmetry $\aslb$, taking advantage of an increased dataset and several improvements to the analysis. These are briefly described in this section.

The largest contribution to the measured uncertainty on $\aslb$ is due to the limited size of the data sample. The D0 detector continues to collect data, with approximately 9~fb$^{-1}$ available for analysis at the time of writing. This represents an increase in integrated luminosity of $\sim$50\% with respect to the published result, therefore an updated measurement using this additional data will significantly increase sensitivity to non-SM effects, even with no analysis improvements.   

The event selection is being revisited, and a few minor changes have been proposed to optimise the signal sample, while maintaining a small background contamination. A new method (the so-called `null fit' descibed in Appendix E of Ref.~\cite{asl-d0}) has been developed to measure the kaon fractions with greater precision, reducing sensitivity to the detector mass resolution and peaking backgrounds in the $K^{*0}$ invariant mass distributions. This new method also enables a second channel to be included in the $F_k/f_k$ measurement, with consistent results obtained for both.

Finally, muon impact parameter (IP) information is being used to help identify the likely source of the observed asymmetry. For muons with small IP (less than 100$\mu$m, for instance), the probability that they arise from a $B_d^0$ meson which has mixed is small, due to the low mixing frequency associated with the mass difference $\Delta M_d \approx 0.5$~ps$^{-1}$. In contrast, the contribution of muons from decays of $B_s^0$ mesons which have mixed prior to decay is relatively stable with IP, due to the much higher mixing frequency associated with the mass difference $\Delta M_s \approx 17$~ps$^{-1}$. As such, by dividing the sample into two parts corresponding to small and large impact parameter muons, the resulting independent measurements of $\aslb$ yield two bands in the $\asld$-$\asls$ plane with different slopes. The overlapping region of these two slopes can then be directly compared to the SM predictions for both $\asld$ and $\asls$.

The anticipated timescale for publication is summer 2011.


\begin{thebibliography}{}
\bibitem{sakharov} A.D.~Sakharov, Pisma Zh.~Eksp.~Teor.~Fiz. {\bf 5}, 32 (1967)
     [Sov.~Phys.~JETP Lett. {\bf 5}, 24 (1967)].
\bibitem{pdg} C.~Amsler {\sl et al.}, Phys.~Lett.~B~{\bf 667}, 1 (2008).
\bibitem{gavela_huet} M.B.~Gavela, P.~Hernandez, J.~Orloff, O.~Pene, Mod. Phys.
     Lett.~A {\bf 9}, 795 (1994); M.B.~Gavela, P.~Hernandez, J.~Orloff, O.~Pene, C.~Quimbay, 
     Nucl. Phys.~B {\bf 430}, 382 (1994); P.~Huet and E.~Sather, Phys. Rev.~D {\bf 51}, 379 (1995).
\bibitem{Nierste} A.~Lenz and U.~Nierste, J.~High Energy Phys.~{\bf0706}, 072 (2007).
\bibitem{Randall} L.~Randall and S.~Su, Nucl.~Phys.~B~\textbf{540}, 37 (1999).
\bibitem{Hewett} J.L.~Hewett, arXiv:hep-ph/9803370 (1998).
\bibitem{Hou} G.W.S.~Hou, arXiv:0810.3396 [hep-ph] (2008).
\bibitem{Soni} A.~Soni {\it et al.}, Phys. Lett. B {\bf 683}, 302 (2010); 
     A.~Soni {\it et al.}, arXiv:1002.0595 (2010) [hep-ph] and references therein.
\bibitem{Buras} M.~Blanke, A.~J.~Buras, A.~Poschenrieder, C.~Tarantino, S.~Uhlig and A.~Weiler,
     JHEP {\bf 0612}, 003 (2006).  W.~Altmannshofer, A.~J.~Buras, S.~Gori, P.~Paradisi and D.~M.~Straub, 
     Nucl.\ Phys.\  B {\bf 830}, 17 (2010).
\bibitem{d0det} V.M.~Abazov {\sl et al.} (D0 Collaboration),
     Nucl. Instrum. Methods Phys. Res. A~{\bf 565}, 463  (2006).
\bibitem{d0asym2mu} V.A.~Abazov {\sl et al.} (D0 Collaboration),
     Phys.~Rev.~D.~\textbf{82}, 032001 (2010).
\bibitem{D01} V.M.~Abazov, \textit{et al.} (D0 Collaboration), Phys.~Rev.~D~\textbf{74}, 092001 (2006).
\bibitem{babar} B.~Aubert \textit{et al.} (Babar Collaboration), Phys.~Rev.~Lett.~\textbf{96}, 251802 (2006); 
     B.~Aubert \textit{ et al.} (Babar Collaboration), arXiv:hep-ex/0607091 (2006).
\bibitem{belle} E.~Nakano \textit{et al.} (Belle Collaboration), Phys.~Rev.~D~\textbf{73}, 112002 (2006).
\bibitem{hfag} E.~Barberio {\it et al.} (HFAG), arXiv:0808.1297 [hep-ex] (2008).
\bibitem{asl-d0} V.M.~Abazov \textit{et al.} (D0 Collaboration), Phys.~Rev.~D~\textbf{82}, 012003 (2010).
\end{thebibliography}
\end{document}